\documentclass[fleqn]{article}
\usepackage{amscd,amsmath,amsthm,amssymb}
\usepackage[dvips]{graphicx}

\theoremstyle{plain}
\newtheorem{thm}{Theorem}[section]
\newtheorem{prop}[thm]{Proposition}
\renewcommand{\abstract}{Abstract.}
\def\H{\mathcal{H}_3}
\def\R{\mathbb{R}}
\def\va{\vec{a}}
\def\vb{\vec{b}}
\def\vc{\vec{c}}
\def\vd{\vec{d}}
\def\eee{\end{document}}
\title{Additive angles in $\H$}
\author{D.G. Pavlov\thanks{Institute of Hypercomplex Geometry, Russian Federation, E-mail: geom2004@mail.ru},
    S.S. Kokarev\thanks{RSEC "Logos", Yaroslavl, Russian Federation, E-mail: logos-distant@mail.ru}}
\date{}
    \begin{document}\maketitle\begin{abstract}
Within the framework of Berwald-Moor Geometry in $\H$, the paper studies the construction of
    additive poly-angles ({\em bingles} and {\em tringles}). It is shown that, considering
    additiveness in the large, there exist an infinity of such poly-angles.
    \end{abstract}
    \section{Introduction}
    Two classical basic notions of Euclidean geometry are {\em the length} and {\em the angle}.
    Both are defined by means of the fundamental metric form --- the scalar product of vectors $\eta:V\times V\to\R$,
    which will be further denoted as $\eta(\va,\vb)$, for all the pairs of vectors $\va$ and $\vb$.
    We denote here by $V$ a finite-dimensional vector space.
    The Euclidean scalar product satisfies three characteristic properties:
    \begin{enumerate}\item $\eta(\va,\vb)=\eta(\vb,\va)$ (symmetry);
    \item $\eta(\lambda\va+\mu\vb,\vc)=\lambda\eta(\va,\vc)+\mu\eta(\vb,\vc)$ (bi-linearity);
    \item $\eta(\va,\va)\ge0$ ($=0$, only for $\va=0$) (positiveness),\end{enumerate}
    where we denoted by $\va,\vb,\vc$ three arbitrary vectors, and by $\lambda,\mu$ --- two
    arbitrary real numbers. Such a scalar product allows to correctly define the length (the norm) $|\va|$ of a vector
    $\va$, and the angle $\varphi(\va,\vb)$ between two nonzero vectors $\va,\vb$, as follows:
        \begin{equation}\label{meas}|\va|=\sqrt{\eta(\va,\va)};\quad
        \varphi(\va,\vb)=\arccos\frac{\eta(\va,\vb)}{|\va||\vb|}.\end{equation}
    The scalar product axioms lead to the following usual properties of these two measures:
    \begin{enumerate}\item
    both measures are invariant w.r.t. the group of motions of the Euclidean space (which are generated by rotations,
    translations and symmetries);\item
    the angle is invariant w.r.t. the action of the large group of conformal symmetries of the Euclidean space,
    which, besides the elements of the motions group, includes as well, as generators, the homogeneous scalings
    of the coordinate axes;\item
    the norm satisfies the triangle inequality: $|\va|+|\vb|\ge|\va+\vb|$, where the equality
        holds true only in the case of collinear vectors having the same orientation, and the equality
        represents the principle of additivity of the Euclidean length of segments;
    \item the positivity property of the scalar product allows to correctly define the cosine of the angle, since:
        \[\left|\frac{\eta(\va,\vb)}{|\va||\vb|}\right|\le1;\]
\item    as well, the additivity of angles takes place:
     \begin{equation}\label{add1}\varphi(\va,\vb)=\varphi(\va,\vc)+\varphi(\vc,\vb),\end{equation}
     for any triple $\{\va,\vb,\vc\}$ of vectors for which
     \begin{equation}\label{comp}\text{vol}(\va,\vb,\vc)=0,\end{equation}
     where  vol is the volume form (the mixed product), with the angles considered as being oriented and, say,
     with the orientation from $\va$ to $\vb$ considered as positive;\item
    the measure of the angle is symmetric w.r.t. the couple of vectors, i.e.,
\begin{equation}\label{symm1}\varphi(\va,\vb)=\varphi(\vb,\va).\end{equation}\end{enumerate}

    The purpose of the present paper is to introduce the Finslerian extension of the notion of angle within
    the framework of the 3-dimensional Berwald-Moor space $\H$, which naturally appears while studying the
    commutative-associative algebras \cite{pavlov}. The basic geometric object of the space $\H$ is the
    Berwald-Moor metric
    \begin{equation}\label{bm}{}^3G=\hat{\mathcal{S}}(dx^1\otimes dx^2\otimes dx^3)\end{equation}
    where $\hat{\mathcal{S}}$ is the symmetrization operator (up to a non-zero real multiplier).
    Based on the conformal invariance and additivity, we shall study the possibility of constructing angles in $\H$.
    Our study will show that the geometry of angles in $\H$ is, in a certain sense, richer than the geometry of
    angles of the Euclidean space. We shall obtain both the exact expressions of some general  types of angles in $\H$, and
    the corresponding group of their symmetries --- which will be shown to be richer than the conformal group
    of symmetries associated to the Euclidean metric.
\section{Additive angles in Euclidean and \protect{\\}pseudo-Euclidean geometries}
    Before studying angles in $\H$, we shall point out the general idea of this research, using the example
    provided by the simple question of finding all the additive angles in
    Euclidean geometry. In other words, we shall forget for the time being the second definition (\ref{meas})
    and we shall try to obtain it (or its extension) by means of only the before mentioned properties of conformal
    invariance and additivity. To achieve this goal, we first remark that for any couple of nonzero vectors
    $\va,\vb$ of the $n-$dimensional Euclidean space, there exist only two functionally independent conformally-invariant
    combinations related to the Euclidean form, namely:
        \begin{equation}\label{base1}w^a_b\equiv\frac{\eta(\va,\vb)}{\eta(\va,\va)}\quad\text{ш}
        \quad w^b_a\equiv\frac{\eta(\va,\vb)}{\eta(\vb,\vb)}.\end{equation}
    The desired angle should be defined as some smooth function of these variables:
        \[\varphi(\va,\vb)=f(w^a_b,w^b_a).\]
    For any third nonzero vector $\vc$, which satisfies the coplanarity condition (\ref{comp}), the condition of
    additivity of angles has the form
        \begin{equation}\label{add2}f(w^a_b,w^b_a)=f(w^a_c,w^c_a)+f(w^c_b,w^b_c).\end{equation}
    To write this condition in the language of local coordinates, and then in the language of functional
    equations in the space of functions of numerical variable, we choose a Cartesian system of coordinates,
    such that: $\va=(a,0,0,\dots,0)$, $\vb=(b_1,b_2,0,\dots,0)$. The coplanarity condition, written for the
    vectors $\va,\vb,\vc$, is equivalent to
    \[\vc= \alpha_1\va+\alpha_2\vb=(\alpha_1a+\alpha_2b_1,\alpha_2 b_2,0,\dots,0),\]
    and hence, from now on, we deal with plane Euclidean geometry. Simple calculations performed with conformally
    invariant combinations from (\ref{add2}) lead to
      \[w^a_b=\frac{b_1}{a};\quad w^b_a=\frac{ab_1}{b_1^2+b_2^2};\quad w^a_c=\alpha_1+\alpha_2\frac{b_1}{a};
      \quad w^c_a=\frac{(\alpha_1a+\alpha_2b_1)a}{(\alpha_1a+\alpha_2b_1)^2+(\alpha_2b_2)^2};\]
      \[w^c_b=\frac{(\alpha_1a+\alpha_2b_1)b_1+\alpha_2b_2^2}{(\alpha_1a+\alpha_2b_1)^2+(\alpha_2b_2)^2};
      \quad w^b_c=\frac{(\alpha_1a+\alpha_2b_1)b_1+\alpha_2b_2^2}{b_1^2+b_2^2}.\]
    Substituting this in (\ref{add2}) and denoting $b_1/a=\xi$, $b_2/a=\eta$, we obtain the following functional
    equation, which expresses the condition of additivity of angles:
        \begin{equation}\label{fund1}\begin{array}{ll}
        f\left(\xi,\dfrac{\xi}{\xi^2+\eta^2}\right)=&f\left(\alpha_1+\alpha_2\xi,
        \dfrac{\alpha_1+\alpha_2\xi}{(\alpha_1+\alpha_2\xi)^2+(\alpha_2\eta)^2}\right)\medskip\\
        &+f\left(\dfrac{(\alpha_1+\alpha_2\xi)\xi+\alpha_2\eta^2}{(\alpha_1+\alpha_2\xi)^2+(\alpha_2\eta)^2},
        \dfrac{(\alpha_1+\alpha_2\xi)\xi+\alpha_2\eta^2}{\xi^2+\eta^2}\right).\end{array}\end{equation}
    Obviously, we can check straightforward, using massive substitutions and considering the
    identity:
    $\arccos x+\arccos y=\arccos[xy-\sqrt{1-x^2}\sqrt{1-y^2}]$, that the equation (\ref{fund1}) admits a solution of
    the form: $f(w_1,w_2)=\arccos{\sqrt{w_1w_2}}$, which corresponds to the standard definition of angle (\ref{meas}).
    To illustrate the fact that our posed problem is non-trivial, we shall subsequently analyze the equation
    (\ref{fund1}). Since this equation has to be satisfied for all the values of the parameters
    $\eta,\xi,\alpha_1,\alpha_2$, it should be satisfied as well on certain submanifolds of the space of these
    parameters. We examine the equation constrained to the submanifold $\eta=0$, i.e., ${\vec a}||{\vec b}$:
        \begin{equation}\label{eq1}f(\xi,1/\xi)=f(\alpha_1+\alpha_2\xi,(\alpha_1+\alpha_2\xi)^{-1})+
        f\left(\frac{\xi}{\alpha_1+\alpha_2\xi},\frac{\alpha_1+\alpha_2\xi}{\xi}\right).\end{equation}
    Denoting $\alpha_1+\alpha_2\xi=x$, the equation (\ref{eq1}) can be written as
       \begin{equation}\label{eq2}f(\xi,1/\xi)=f(x,1/x)+f(\xi/x,x/\xi).\end{equation}
    The equation (\ref{eq2}) has to be satisfied for all the values of the variables $\xi,x$.
    Denoting $f(w,1/w)\equiv F(w)$,, we get the equivalent equation
       \begin{equation}\label{eq3}F(\xi)=F(x)+F(\xi/x),\end{equation}
    which has to be satisfied for all $\xi,x$ as well.
    Assuming the smoothness of the function $F$, and deriving the equation (\ref{eq3}) w.r.t. $\xi$ and $x$, we yield
        \[\frac{\partial^2 F(\xi/x)}{\partial x\partial\xi}=0\]
    or
        \begin{equation}\label{eq4}F^{\prime\prime}u+F^{\prime}=0,\end{equation}
    where $u=\xi/x$, and prime denotes the derivative w.r.t. the whole argument. The equation (\ref{eq4}) has the
    general solution of the form
       \begin{equation}\label{sol1}F(u)=C_1\ln u+C_2,\end{equation}
    where $C_1, C_2$ are arbitrary constants. We show that the solution $F=C_2=\text{const}$ corresponds to the
    standard definition of angle (\ref{meas}). Indeed, recollecting that $F(x)=f(x,1/x)$, we conclude that the
    mapping $f(w_1,w_2)$ is constant for $w_1=x$ and $w_2=1/x$ for all the values of $x$, only when it depends
    on the product $w_1\cdot w_2$, i.e.,
       \begin{equation}\label{sol2}f(w_1,w_2)=\psi(w_1w_2).\end{equation}
    Then, making use of (\ref{eq2}), we infer $\psi(1)=0$. Now we come back to the equation (\ref{fund1}).
    Using in it the form (\ref{sol2}), and examining the obtained equation on the submanifold $\alpha_1=\alpha_2$,
    we infer the equation
        \begin{equation}\label{eq5}\psi\left(\frac{\xi^2}{\xi^2+\eta^2}\right)=
        \psi\left(\frac{(1+\xi)^2}{(1+\xi)^2+\eta^2}\right)+\psi\left(\frac{((1+\xi)\xi+\eta^2)^2}
        {((1+\xi)^2+\eta^2)(\xi^2+\eta^2)}\right).\end{equation}
    We pass to the new variables
        \[u=\frac{\xi^2}{\xi^2+\eta^2};\quad v=\frac{(1+\xi)^2}{(1+\xi)^2+\eta^2}.\]
    With respect to these variables, the equation (\ref{eq5}) gets a simpler form:
        \begin{equation}\label{eq6}\psi(u)=\psi(v)+\psi((\sqrt{(v-1)(u-1)}+\sqrt{uv})^2).\end{equation}
    Deriving this equation subsequently relative to $u$ and $v$, we get the consequence of (\ref{eq6}):
        \[\frac{\partial^2}{\partial v\partial u}\;\;\psi((\sqrt{(v-1)(u-1)}+\sqrt{uv})^2)=0\]
    and, after several elementary transformations, we obtain the differential equation:
        \begin{equation}\label{eq7}(1-\zeta^2)\bar\psi^{\prime\prime}(\zeta)-\zeta\bar\psi'(\zeta)=0,\end{equation}
    where $\zeta=\sqrt{(v-1)(u-1)}+\sqrt{uv}$ and $\bar\psi(x)\equiv\psi(x^2)$. We integrate the equation
    (\ref{eq7}); its general solution has the form:
        \begin{equation}\label{sol3}\bar\psi(\zeta)=A\arcsin\zeta+B\Rightarrow\psi(x)=
        A\arcsin\sqrt{x}+B.\end{equation}
    Considering the condition $\psi(1)=0$, we get $B=-\pi A/2$, whence we infer the solution in its final form:
        \[\psi(x)=A\cdot(\arcsin\sqrt{x}-\pi/2)=A\arccos\sqrt{x},\]
    which, up to the choice of unity of the measure of angle, given by the parameter $A$, is equivalent to the
    second definition (\ref{meas}). We shall examine now the second independent solution, which in (\ref{sol1})
    corresponds to the logarithm. We study the general equation (\ref{fund1}) on the submanifold $\alpha_1=0:$
        \begin{equation}\label{eq8}f\left(\xi,\frac{\xi}{\xi^2+\eta^2}\right)=f\left(\alpha_2\xi,
        \frac{\xi}{\alpha_2(\xi^2+\eta^2)}\right)+f\left(\dfrac{1}{\alpha_2},\alpha_2\right).\end{equation}
    If we additionally impose on the variables the submanifold constraint: $\xi/(\xi^2+\eta^2)=k/\xi$, where
    $k=\text{const}$, and consider that for the examined branch of solutions --- as it was established before,
    we have $f(x,1/x)=C\ln x$, then the relation (\ref{eq8}) gets the form:
        \[f(\xi,k/\xi)=f(\alpha_2\xi,k/\alpha_2\xi)+C\ln\alpha_2.\]
    Further, denoting $\Phi_k(\xi)\equiv f(\xi,k/\xi)$, the last equation can be written as:
        \[\Phi_k(\alpha\xi)=\Phi_k(\xi)-C\ln\alpha,\]
    for all the values of $\xi$, $\alpha$ and $k$. We can analogously to (\ref{eq3}) solve the last equation, and its
    general solution has the form:
        \[\Phi_k(\xi)=\ln(\xi^{-C})+B_k,\]
    where $B_k$ is an arbitrary function of $k$.
    Reminding that $\Phi_k(\xi)=f(\xi,k/\xi)$ and comparing this with the obtained solution, we infer that the
    function $f(w_1,w_2)$ may depend on its arguments as follows (where we omit the additive constant, since it
    corresponds to the studied arccosine branch):
        \begin{equation}\label{sol4}f(w_1,w_2)=\ln w_1^Aw_2^B+\varphi(w_1w_2) ,\end{equation}
    where $A,B$ are (for the time being) arbitrary real numbers, and $\varphi$ is an arbitrary function, which
    describes the studied branch of the solution. To specify the values of the constants, we replace the
    solution (\ref{sol4}) (dropping the addition with $\varphi$) in the general equation (\ref{fund1}).
    As consequence, we get an equation
    which is satisfied for all the values of the variables, only if $A=-B$. Then $f(w_1,w_2)=A\ln(w_1/w_2)$.
    The results obtained above can be summarized as follows.\par\medskip
\begin{prop}\label{pu}
    There exists a pair of functionally independent conformally-invariant additive expressions for the angle between
    the vectors $\va$ and $\vb$ in Euclidean geometry, namely:
        \begin{equation}\label{solg1}\varphi_1(\va,\vb)=\arccos\left(\frac{\eta(\va,\vb)}{|\va||
        \vb|}\right);\quad \varphi_2(\va,\vb)=\ln\left(\frac{|\va|}{|\vb|}\right).\end{equation}
    We have:
        \[\varphi_1(\va,\vb)=\varphi_1(\vb,\va);\quad \varphi_2(\va,\vb)=-\varphi_2(\vb,\va)\]
    and the general expression for "the additive conformally-invariant angle"\, in Euclidean geometry is given by
    the combination of the two independent expressions:
        \[\varphi(\va,\vb)=C_1\varphi_1(\va,\vb)+C_2\varphi_2(\va,\vb),\]
        where $C_1$ and $C_2$ are arbitrary constants.
\end{prop}
    We note, that similar reasoning as in Proposition \ref{pu} leads to the corresponding result for the
    pseudo-Euclidean plane:
\begin{prop}\label{pd}There exists a pair of linearly independent conformally-invariant additive expressions for the
    angle between the vectors $\va$ ш $\vb$ in pseudo-Euclidean geometry (where $\eta$ is the pseudo-Euclidean metric
    in plane):
        \begin{equation}\label{solg2}\varphi_1(\va,\vb)={\rm arccosh}\left(\frac{\eta(\va,\vb)}
        {|\va||\vb|}\right);\quad \varphi_2(\va,\vb)=\ln\left(\frac{|\va|}{|\vb|}\right).\end{equation}
        We have:
        \[\varphi_1(\va,\vb)=\varphi_1(\vb,\va);\quad \varphi_2(\va,\vb)=-\varphi_2(\vb,\va)\]
    and the general expression for "the additive conformally-invariant angle"\, in pseudo-Euclidean geometry is
    given by the combination of the independent expressions:
        \[\varphi(\va,\vb)=C_1\varphi_1(\va,\vb)+C_2\varphi_2(\va,\vb),\]
        where $C_1$ and $C_2$ are arbitrary real constants.
\end{prop}
    Strictly speaking, while passing to the formulas (\ref{sol2}) and (\ref{sol4}), we did not examine the most
    general case, and hence the statement regarding the generality of the expressions (\ref{solg1}) and (\ref{solg2})
    needs supplementary considerations. Namely, the adding of a term of the form $(w_1w_2-1)\varphi(w_1,w_2)$ to the
    functions (\ref{sol2}) and (\ref{sol4}), where $\varphi(w_1,w_2)$ is finite for $w_1w_2=1$, does not change the
    conditions which have been earlier derived for these functions, since this additional term vanishes for
    $w_1w_2=1$. We shall further prove, that $\varphi\equiv0$. Replacing $f=(w_1w_2-1)\varphi(w_1,w_2)$ in the general
    equation (\ref{fund1}), after several simplifying calculations, we get:
        \begin{equation}\label{fund11}\begin{array}{c}\frac{\eta^2}{\xi^2+\eta^2}\varphi\left(\xi,\frac{\xi}
        {\xi^2+\eta^2}\right)=\frac{(\alpha_2\eta)^2}{(\alpha_1+\alpha_2\xi)^2+(\alpha_2\eta)^2}\cdot\varphi
        \left(\alpha_1+\alpha_2\xi,\frac{\alpha_1+\alpha_2\xi}{(\alpha_1+\alpha_2\xi)^2+(\alpha_2\eta)^2}\right)
        \medskip\\
        +\frac{(\alpha_1\eta)^2}{((\alpha_1+\alpha_2\xi)^2+(\alpha_2\eta)^2)(\xi^2+\eta^2)}\cdot\varphi
        \left(\frac{(\alpha_1+\alpha_2\xi)\xi+\alpha_2\eta^2}{(\alpha_1+\alpha_2\xi)^2+(\alpha_2\eta)^2},
        \frac{(\alpha_1+\alpha_2\xi)\xi+\alpha_2\eta^2}{\xi^2+\eta^2}\right).\end{array}\end{equation}
    We shall first examine this equation on the submanifold $\alpha_1=0$. After dividing the equation by the common
    multiplier, we get:
        \[\varphi\left(\xi,\frac{\xi}{\xi^2+\eta^2}\right)=\varphi\left(\alpha_2\xi,
        \frac{\xi}{\alpha_2(\xi^2+\eta^2)}\right),\]
    whence it follows that the function $\varphi$ may depend on its arguments only by means of their product, i.e.:
    $\varphi=\varphi(w_1w_2)$. Replacing this expression in the equation (\ref{fund11}) and examining this equation
    on the submanifold $\xi=0$ we infer, after the change of variables, the equation:
        \[\varphi(0)=y\varphi(x)+x\varphi(y).\]
    This equation is identically satisfied for all $x$ and $y$ only for $\varphi\equiv0$, q.e.d..\par
    The expressions (\ref{solg1}) and (\ref{solg2}) represent indeed the most general expressions of additive
    angles on the Euclidean and pseudo-Euclidean planes, respectively. In this case, we can easily note that the
    additivity $\varphi_2(\va,\vb)$ is unconditionally satisfied, i.e., it is satisfied for any arbitrary triple
    of vectors, not only for coplanar vectors.
\section{The additive bingle in $\H$ --- the affine version}\label{affinn}
    In the Berwald-Moor space, there exist two natural ways of introducing angles, namely: the ones built on two vectors
    (bingles) and the ones built on triples of vectors (tringles). The latter ones represent essentially new objects,
    which are tightly related to the cubic character of the metric (\ref{bm}) in $\H$, and which have no direct
    analogues in Euclidean geometry. In this Section we aim to systematically determine the bingles in $\H$.\par
    By analogy to the Euclidean case we shall examine the elementary conformally-invariant combinations which are
    built on pairs of vectors $\va$ and $\vb$ by means of the Berwald-Moor metric ${}^3G$:
        \[w^a_a\equiv\frac{{}^3G(\va,\va,\vb)}{{}^3G(\va,\va,\va)};\quad
        w^a_b\equiv\frac{{}^3G(\va,\va,\vb)}{{}^3G(\vb,\vb,\vb)};\quad
        w^b_a\equiv\frac{{}^3G(\va,\vb,\vb)}{{}^3G(\va,\va,\va)};\quad
        w^b_b\equiv\frac{{}^3G(\va,\vb,\vb)}{{}^3G(\vb,\vb,\vb)}.\]
        Using the vector coordinates $\va=(a_1,a_2,a_3)$, $\vb=(b_1,b_2,b_3)$ and denoting $\xi_i\equiv b_i/a_i$
        $(i=1,2,3)$, we get:
        \[\begin{array}{l}w^a_a=\frac{1}{3}(\xi_1+\xi_2+\xi_3);\quad w^a_b=\frac{1}{3}\left(\frac{1}{\xi_1\xi_2}+
        \frac{1}{\xi_1\xi_3}+\frac{1}{\xi_2\xi_3}\right);\medskip\\
        w^b_a=\frac{1}{3}(\xi_1\xi_2+\xi_1\xi_3+\xi_2\xi_3);\quad
        w^b_b=\frac{1}{3}\left(\frac{1}{\xi_1}+\frac{1}{\xi_2}+\frac{1}{\xi_3}\right).\end{array}\]
    Taking into account that
        \[\frac{w^a_a}{w^a_b}=\xi_1\xi_2\xi_3\quad\text{ш}\quad w^b_b=\frac{w^b_a w^a_b}{w^a_a},\]
    we can consider as independent conformally-invariant variables, three symmetric polynomials:
        \begin{equation}\label{invar}w^{ab}_1=\xi_1+\xi_2+\xi_3;\quad
        w^{ab}_2=\xi_1\xi_2+\xi_1\xi_3+\xi_2\xi_3;\quad w^{ab}_3=\xi_1\xi_2\xi_3, \end{equation}
    where $\varphi(\va,\vb)=f(w^{ab}_1,w^{ab}_2,w^{ab}_3)$ is the angle, with $f$ a function which has to be
    determined.\par
    By analogy to the Euclidean case, the property of additivity of angles in the space $\H$ should be stated
    not for an arbitrary triples of vectors, but on triples which satisfy certain relations. In Euclidean and
    pseudo-Euclidean geometry, such a relation is the {\em coplanarity}. But in the $\H-$geometry, we cannot expect
    the plane to play the same role like in the geometries with quadratic metric. Still, in this section we shall
    investigate the possibility of defining additive vectors for triples of vectors which are coplanar in the usual
    sense. This case will be called {\em the affine version of the additive angle}.
    Let the third vector $\vc$ belong to the plane determined by the vectors $\va$ and $\vb$:
        \begin{equation}\label{vecc}\vc=\alpha_1\va+\alpha_2\vb=(\alpha_1a_1+\alpha_2b_1,\alpha_1a_2+
        \alpha_2b_2,\alpha_1a_3+\alpha_2b_3).\end{equation}
    Then the additivity condition takes the form:
        \begin{equation}\label{Add2}f(w^{ab}_1,w^{ab}_2,w^{ab}_3)=f(w^{ac}_1,w^{ac}_2,w^{ac}_3)+
        f(w^{cb}_1,w^{cb}_2,w^{cb}_3),\end{equation}
    which should hold true for all the (non-isotropic) vectors $\va,\vb$ and for any vector $\vc$ of the form
    (\ref{vecc}). Computing $w^{ac}_i$ and $w^{cb}_i$ by means of (\ref{vecc}) and substituting the result into
    (\ref{Add2}), we obtain, after a simplifying process, the following functional equation:
        \begin{equation}\label{Add21}\begin{array}{l}
        f(w_1,w_2,w_3)=f(3\alpha_1+\alpha_2w_1,3\alpha_1^2+\alpha_2^2w_2+
        2\alpha_1\alpha_2w_1,\Delta)\medskip\\
        +f\left(\dfrac{\alpha_1^2w_1+3\alpha_2^2w_3+2\alpha_1\alpha_2w_2}{\Delta},
        \dfrac{\alpha_1w_2+3\alpha_2w_3}{\Delta},\dfrac{w_3}{\Delta}\right),\end{array}\end{equation}
    which explicitly expresses in coordinates the condition of additivity of angles. Here we have denoted
        \begin{equation}\label{star}\Delta=\alpha_1^3+\alpha_2^3w_3+\alpha_1^2\alpha_2 w_1+\alpha_1\alpha_2^2w_2.
        \end{equation}
    We shall examine the relation (\ref{Add21}), which, as before, should hold true for all the values of the five
    variables $w_1,w_2,w_3,\alpha_1,\alpha_2$, on the 2-dimensional submanifold defined by the relations:
        \[w_1=-\frac{3\alpha_1}{\alpha_2};\quad w_2=\frac{3\alpha_1^2}{\alpha_2^2};\quad
        w_3=\frac{1-\alpha_1^3}{\alpha_2^3}.\]
    On this submanifold the equation (\ref{Add21}) has the simple form:
        \[f\left(-\frac{3\alpha_1}{\alpha_2},\frac{3\alpha_1^2}{\alpha_2^2},\frac{1-\alpha_1^3}{\alpha_2^3}\right)=
        f(0,0,1)+f\left(\frac{3}{\alpha_2},\frac{3}{\alpha_2^2},\frac{1-\alpha_1^3}{\alpha_2^3}\right)\]
    or, using the variables $x,y$ where $\alpha_1=x,\alpha_2=1/y$, we get
        \[f(-3xy,3x^2y^2,(1-x^3)y^3)=f(0,0,1)+f(3y,3y^2,(1-x^3)y^3).\]
    For $y=0$, we obtain the equality $f(0,0,1)=0$, and hence, in fact, we deal with the equation:
        \[f(-3xy,3x^2y^2,(1-x^3)y^3)=f(3y,3y^2,(1-x^3)y^3).\]
    Differentiating both sides of this equation by $x$ and further by replacing $x=-1$, we get, after renoting the
    variables, the PDE of first order:
        \[x_1\frac{\partial f(x_1,x_2,x_3)}{\partial x_1}+2x_2\frac{\partial f(x_1,x_2,x_3)}{\partial x_2}=0.\]
    We integrate it by the method of characteristics, and we infer:
        \begin{equation}\label{solb1}f(w_1,w_2,w_3)=\psi(w^2_1/w_2,w_3),\end{equation}
    where $\psi$ is some function, which depends only on two variables.
    Replacing this into the original equation (\ref{Add21}) and examining the resulting equation on the
    submanifold $\alpha_1=0$, we obtain:
        \[\psi\left(\frac{w_1^2}{w_2},w_3\right)=\psi\left(\frac{w_1^2}{w_2},\alpha_2^3w_3\right)+\psi
        \left(3;\frac{1}{\alpha_2^3}\right).\]
    From this relation it follows that, w.r.t. the second argument, the dependence is of logarithmic nature, and
    hence:
        \[\psi(w_1^2/w_2,w_3)=\chi(w_1^2/w_2)\ln w_3+C(w_1^2/w_2).\]
    In order to determine the remaining functions $\chi$ and $C$, we replace this expression into the previous
    relation. We finally get:
        \begin{equation}\label{solbh3}C=0,\quad \chi=\text{const}.\end{equation}\par
    To have the complete picture, it remains to investigate the case when the additive angle is assumed to depend
    not on all the variables $w_1,w_2,w_3$, but only on a part of them.\par\smallskip
{\bf I a.} We first assume that
    ${}^3\varphi(\va,\vb)=f(w_1,w_2)$. Using the general formula (\ref{Add21}), we obtain in this case the
    additivity condition of the form:
        \begin{equation}\label{Add22}\begin{array}{ll}
        f(w_1,w_2)=&f(3\alpha_1+\alpha_2w_1,3\alpha_1^2+\alpha_2^2w_2+
        2\alpha_1\alpha_2w_1)\medskip\\
        &+f\left(\dfrac{\alpha_1^2w_1+3\alpha_2^2w_3+2\alpha_1\alpha_2w_2}{\Delta},
        \dfrac{\alpha_1w_2+3\alpha_2w_3}{\Delta}\right),\end{array}\end{equation}
    with $\Delta$ defined in (\ref{star}).
    We partially derive this relation w.r.t. $w_3$, and we infer the equation:
        \[\frac{\partial f}{\partial u_1}(3\alpha_1^2\Delta-(\alpha_1^2w_1+3\alpha_2^2w_3+
        2\alpha_1\alpha_2w_2)\alpha_2^3)+\frac{\partial f}{\partial u_2}(3\alpha_2\Delta-
        (\alpha_1w_2+3\alpha_2w_3)\alpha_2^3)=0,\]
    where $u_1$ and $u_2$ are the arguments of the function $f$ in the second term of the right side of
    (\ref{Add22}).
    We shall examine this equation on the submanifold $\alpha_1=\alpha_2=\alpha$; we yield:
        \[\begin{array}{c}
        \alpha\dfrac{\partial f}{\partial u_1}\left(\dfrac{w_1+3w_3+2w_2}{\alpha(1+w_1+w_2+w_3)},\dfrac{w_2+
        3w_3}{\alpha(1+w_1+w_2+w_3)}\right)(3+2w_1+w_2)\medskip\\
        +\dfrac{\partial f}{\partial u_2}\left(\dfrac{w_1+3w_3+2w_2}{\alpha(1+w_1+w_2+w_3)},
        \dfrac{w_2+3w_3}{\alpha(1+w_1+w_2+w_3)}\right)(3+3w_1+2w_2)=0.\end{array}\]
    Passing to the new variables $u_1$ and $u_2$, we write this equation in the form:
        \[\alpha(u_1\alpha-3)\frac{\partial f}{\partial u_1}+(u_2\alpha-3)\frac{\partial f}{\partial u_2}=0.\]
    After solving it by means of the method of characteristics, we infer
        \[f(u_1,u_2)=F\left(\frac{u_1-3/\alpha}{u_2-3/\alpha^2}\right).\]
    But the universal function of angle cannot depend on the coordinates $\alpha_1$ and $\alpha_2$ of the third
    arbitrary vector. Hence we infer the conclusion that $f=\text{const}$.\par\smallskip
{\bf I b.} Assume now that ${}^3\varphi(\va,\vb)=f(w_1,w_3)$. Using the general formula (\ref{Add21}), we get in this
    case the condition of additivity of the form:
        \begin{equation}\label{Add23}f(w_1,w_3)=f(3\alpha_1+\alpha_2w_1,\Delta)+f\left(\frac{\alpha_1^2w_1+
        3\alpha_2^2w_3+2\alpha_1\alpha_2w_2}{\Delta},\frac{w_3}{\Delta}\right).\end{equation}
    We study this equation on the submanifold $3\alpha_1+\alpha_2 w_1=0$, $\Delta=1$. This pair of equations
    practically determine $\alpha_1$ and $\alpha_2$. Extracting from the equations of the submanifolds these
    parameters, substituting them further in the equation (\ref{Add23}) and deriving it partially w.r.t. $w_2$,
    we get an equation of the form:
        \[\frac{\partial f}{\partial u_1}(\dots)=0,\]
    where the parantheses which embed dots denote huge expressions which depend on $w_1,w_2,w_3$ and which are not
    everywhere zero. This implies that $f$ does not depend on $u_1$, and we get the dependence on just one argument,
    case which will be investigated below.\par\smallskip
{\bf I c.} Consider now ${}^3\varphi(\va,\vb)=f(w_2,w_3)$. Using again the general formula (\ref{Add21}), we obtain
    the condition of additivity of the form:
        \begin{equation}\label{Add24}f(w_2,w_3)=f(3\alpha_1+\alpha_2^2w_2+2\alpha_1\alpha_2w_1,\Delta)+
        f\left(\frac{\alpha_1w_2+3\alpha_2w_3}{\Delta},\frac{w_3}{\Delta}\right).\end{equation}
    Applying the submanifold constraint $\Delta=1$, $\alpha_1w_2+3\alpha_2w_3=0$ as above, we infer as well
    ${\partial f}/{\partial u_1}=0$, i.e., the dependence of $f$ on one variable.\par\smallskip
{\bf II a.} Assuming that $f=f(w_1)$, we get the additivity condition in the form:
        \[f(w_1)=f(3\alpha_1+\alpha_2w_1)+f\left(\frac{\alpha_1^2w_1+3\alpha_2^2w_3+
        2\alpha_1\alpha_2w_2}{\Delta}\right),\]
    whence, partially deriving, e.g., by $w_2$, we obtain the equality $f^{\prime}=0$ and hence $f=\text{const}$.
    Moreover, the additivity will hold true only for $f=0$. \par\smallskip
{\bf II b.} Analogously, assuming $f=f(w_2)$, we infer the
    equation of additivity:
        \[f(w_2)=f(3\alpha_1^2+\alpha_2^2w_2+2\alpha_1\alpha_2w_1)+f\left(\frac{\alpha_1w_2+
        3\alpha_2w_3}{\Delta}\right).\]
    Applying the constraint $\alpha_1w_2+3\alpha_2w_3=0$, (which provides, say, $\alpha_1$), and deriving the
    equation w.r.t. $w_2$ we infer $f'=0$ and hence, reasoning as before, $f=0$.\par\smallskip
{\bf II c.} At last, for $f=f(w_3)$, the general condition (\ref{Add22}) leads to the equation
        \[f(w_3)=f(\Delta)+f(w_3/\Delta),\]
    whose general solution is of the form (\ref{sol1}). We conclude, stating the following
\begin{prop}\label{pt}The only additive bingle of $\H$ (in the affine version) is defined by:
        \begin{equation}\label{solg3}{}^3\varphi(\va,\vb)=A\ln |w_3|=
        A\ln\left|\frac{{}^3G(\vb,\vb,\vb)}{{}^3G(\va,\va,\va)}\right|.\end{equation}
\end{prop}
    We remark that the expression (\ref{solg3}) is the Finslerian analogue of the angle $\varphi_2$ which
    has been defined in (\ref{solg1}).
    As well, like in the (pseudo-)Euclidean case, ${}^3\varphi$ is additive, regardless of coplanarity. An essential
    consequence of Proposition \ref{pt} is the absence in $\H$ of an analogue to the additive bingle $\varphi_1$
    defined in (\ref{solg1}).
\section{The additive bingle in $\H$ --- the nonlinear version}
    If we omit the supplementary condition of coplanarity of vectors - for which holds true the additivity of angles,
    then the number of possible extensions considerably increases.
\subsection{The existence of nontrivial additive bingles in $\H$ for the nonlinear version}\label{nontr}
    For proving the existence of nontrivial additive bingles, it suffices to provide an example of such a bingle in
    explicit form.\par
    We shall consider as "generalized coplanarity condition" the following generally nonlinear relation satisfied by
    the intermediate vector $\vc$, relative to the vectors $\va$ and $\vb$:
        \[\vc=\phi(w^{ab}_1,w^{ab}_2,w^{ab}_3)\va,\]
    where $\phi$ is an arbitrary function of three arguments, $w^{ab}_i$ are provided by the formulas (\ref{invar}),
    and, as previously, $\xi_i=b_i/a_i$.\par
    We can directly check that the function
        \begin{equation}\label{examp}{}^3\varphi_{(A,B,C)}(\va,\vb)\equiv\ln
        |(w_1^{ab})^A(w_2^{ab})^B(w_3^{ab})^C|-(A+B)\ln3\end{equation}
    satisfies the additivity condition (\ref{Add2}) and defines a generalized conformally-invariant angle between the vectors
    $\va$ and $\vb$, which depends on the arbitrary real parameters $A,B,C$.\par
    We note that, though the provided example is to a certain extent artificial (the vector $\vc$ has to be
    practically collinear to the first vector), the expression of the angle is nontrivial:
        \begin{enumerate}\item this angle, generally speaking, is not symmetric; namely, we have:
        \[{}^3\varphi_{(A,B,C)}(\vb,\va)={}^3\varphi_{(B,A,-C-A-B)}(\va,\vb);\]
    The symmetry of the angle takes place only if $A=B=-C$;\item the angle between two equal vectors is zero:
        \[{}^3\varphi(\va,\va)=0,\]
    and, if considered between collinear vectors, it is expressed by means of their similarity coefficient:
        \[{}^3\varphi(\va,\vec\lambda a)=(A+2B+3C)\ln\lambda;\]
    Two collinear vectors form a null angle only for $A+2B+3C=0$. In particular, this condition is satisfied
    automatically, if the condition of symmetry of angle is fulfilled.\end{enumerate}
    It is obvious, that the 3-parametric family of angles which has been examined, is a nontrivial Finslerian
    extension of the angle $\varphi_2$ from Euclidean geometry.
\subsection{The additive bingle in $\H$ with additivity of angles for 3-orthogonal vectors}
    In this section we examine as \ "generalized coplanarity condition", the condition of 3-orthogonality of the
    form:
        \begin{equation}\label{compf}{}^3G(\va,\vb,\vc)=0,\end{equation}
    which is intrinsic - i.e., natural for the Berwald-Moor geometry in $\H$. In other words, we look for an angle
    function which should be additive on any triple of vectors which satisfy the condition (\ref{compf}).
    This condition has the coordinate expression:
        \begin{equation}\label{av}c_1(a_2b_3+a_3b_2)+c_2(a_1b_3+a_3b_1)+c_3(a_2b_1+a_1b_2)=0.\end{equation}
    The expression (\ref{av}) has the formal shape of the usual Euclidean orthogonality condition of the vector $\vc$
    to the "vector" $\va\bigcirc\vb$ whose components are: $(a_2b_3+a_3b_2,a_1b_3+a_3b_1,a_2b_1+a_1b_2)$. Taking
    into account the well-known properties of the cross product, the general expression for the vector $\vc$, which
    satisfies the Euclidean orthogonality condition, can be written by means of an arbitrary vector
    $\vec\alpha=(\alpha_1,\alpha_2,\alpha_3)$ and the operation of the Euclidean cross product:
        \[\vc=\vec \alpha\times(\va\bigcirc\vb).\]
    Representing the components of the vector $\vec \alpha$ in the form: $\alpha_i=k_i/a_i$, where $k_i$ are
    dimensionless constants, denoting $\xi_i=b_i/a_i$ $(i=1,\dots,3)$, and computing the components of the cross
    product, we find the explicit expressions of the components of $\vc$, which automatically satisfies the
    3-orthogonality condition w.r.t. the given vectors $\va$ and $\vb$:
        \[\left\{\begin{array}{l}
        c_1=k_2(a_1\xi_2+b_1)-k_3(a_1\xi_3+b_1)\medskip\\
        c_2=k_3(a_2\xi_3+b_2)-k_1(a_2\xi_1+b_2)\medskip\\
        c_3=k_1(a_3\xi_1+b_3)-k_2(a_3\xi_2+b_3).\end{array}\right.\]
    By means of these formulas, the conformal invariants, which appear in (\ref{Add2}), get the form:
        \[\begin{array}{ll}
        w_1^{ac}=&k_1(\xi_3-\xi_2)+k_2(\xi_1-\xi_3)+k_3(\xi_2-\xi_1);\medskip\\
        w_2^{ac}=&-k_1^2(\xi_1+\xi_2)(\xi_1+\xi_3)-k_2^2(\xi_1+\xi_2)(\xi_2+\xi_3)-k_3^2(\xi_2+\xi_3)(\xi_1+
            \xi_3)\medskip\\
        &+2k_1k_2(\xi_1+\xi_2)\xi_3+2k_1k_3(\xi_1+\xi_3)\xi_2+2k_2k_3(\xi_2+\xi_3)\xi_1;\medskip\\
        w_3^{ac}=&k_1k_2(\xi_1+\xi_2)^2(k_2(\xi_2+\xi_3)-k_1(\xi_1+\xi_3))\medskip\\
        &+k_1k_3(\xi_1+\xi_3)^2(k_1(\xi_1+\xi_2)-k_3(\xi_2+\xi_3))\medskip\\
        &+k_2k_3(\xi_2+\xi_3)^2(k_3(\xi_1+\xi_3)-k_2(\xi_1+\xi_2)).\medskip\\
        w_1^{cb}=&\frac{k_1\xi_1^2(\xi_2-\xi_3)+k_2\xi_2^2(\xi_3-\xi_1)+k_3\xi_3^2(\xi_1-
            \xi_2)}{\xi_1\xi_2\xi_3};\medskip\\
        w_2^{cb}=&\frac{2k_1k_2}{\xi_3}(\xi_1+\xi_2)+\frac{2k_1k_3}{\xi_2}(\xi_1+\xi_3)+
            \frac{2k_2k_3}{\xi_1}(\xi_2+\xi_3);\medskip\\
        w_3^{cb}=&\frac{k_1k_2}{\xi_1\xi_2\xi_3}(\xi_1+\xi_2)^2(k_2(\xi_2+\xi_3)-k_1(\xi_1+\xi_3))\medskip\\
        &+\frac{k_1k_3}{\xi_1\xi_2\xi_3}(\xi_1+\xi_3)^2(k_1(\xi_1+\xi_2)-k_3(\xi_2+\xi_3))\medskip\\
        &+\frac{k_2k_3}{\xi_1\xi_2\xi_3}(\xi_2+\xi_3)^2(k_3(\xi_1+\xi_3)-k_2(\xi_1+\xi_2)).\end{array}\]
    Unfortunately, after replacing in the additivity condition (\ref{Add2}) the computed invariants, we get
    extremely huge expressions. In order to simplify the obtained relation, we pass to the submanifold
    $k_1=k_2=k_3=k$. Then the additivity condition as defined in Section \ref{affinn}, gets a symmetric form:
        \begin{equation}\label{addfin}\begin{array}{ll}
        f(w_1,w_2,w_3)=&f(0,k^2(3w_2-w_1^2),-k^3w)\medskip\\
        &+f\left(\frac{kw}{w_3},\frac{2k^2(w_1w_2-3w_3)}{w_3},\frac{-k^3w}{w_3}\right).\end{array}\end{equation}
    where $w=(\xi_1-\xi_2)(\xi_2-\xi_3)(\xi_1-\xi_3)$ and we have used the notations in (\ref{invar}). Further,
    constraining the equation (\ref{addfin}) to the submanifold $\xi_1=\xi_2=\xi_3=\xi$, we get an equation of the
    form:
        \begin{equation}\label{addfin1}f(3\xi,3\xi^2,\xi^3)=f(0,0,0)+f(0,12k^2,0),\end{equation}
    whence firstly it follows that the function $f$ does not depend on its second argument, and secondly, that
    the allowed dependence of $f$ on the remaining two arguments has the form:
    $f(w_1,w_3)=\Phi(w_1^3/w_3)$.\par
    Substituting this to the initial equation (\ref{addfin}), we get the simpler equation:
        \begin{equation}\label{addfin2}\Phi\left(\frac{w_1^3}{w_3}\right)=\Phi(0)+
        \Phi\left(-\frac{w^2}{w_3^2}\right).\end{equation}
    Since $w^2$ is a symmetric polynomial of sixth order, it can be expressed in terms of the polynomials $w_1,w_2,w_3$.
    Simple but tedious calculations lead to the following form of $w^2$:
        \begin{equation}\label{nabla}w^2=-4w_2^3-27w_3^2+w_1^2w_2^2-4w_1^3w_3+18w_1w_2w_3.\end{equation}
    It is obvious, that the equation (\ref{addfin2}) cannot admit $\Phi$ as solution for arbitrary values of
    $w_1,w_2,w_3$, unless we apply a constraint on the space of these variables, of the form $w_2=w_2(w_1,w_3)$;
    this is motivated by the fact that the left side does not depend on $w_2$, and the right one, does. Taking into
    account the homogeneity, such a relation can be written as:
        \[w_2=w_1^2\phi(w_1^3/w_3),\]
    where $\phi$ is an arbitrary function. Replacing this into (\ref{nabla}) and further, replacing the
    obtained result into (\ref{addfin2}), we get the additivity equation:
        \[\Phi(4x^2\phi^3(x)-x^2\phi^2(x)-18x\phi(x)+4x+27)=\Phi(x)-\Phi(0),\]
    where $x=w_1^3/w_3$. Denoting by $\phi(x)$ a solution of the equation:
        \[4x^2\phi^3(x)-x^2\phi^2(x)-18x\phi(x)+4x+27=\psi(x),\]
    the additivity equation can be expressed in the general form:
        \begin{equation}\label{addfin3}(\Phi\circ\psi)(x)=\Phi(x)-\Phi(0),\end{equation}
    which defines, in general, a nonlinear homomorphism of the discrete subgroup of homomorphisms $\psi:\R\to\R$
    of the group of translations in the space of angles.\par
    We shall examine two explicit realizations of this homomorphism. The simplest and most natural form of the
    submanifold is provided by the relation $w_2=\xi_1\xi_2+\xi_1\xi_3+\xi_2\xi_3=0$. In this case, the equation
    (\ref{addfin3}) gets the form:
        \[\Phi(x)=\Phi(0)+\Phi(4x+27),\]
    where $x=w_1^3/w_3$. The solution of this equation is the function:
        \[\Phi(x)=\ln\left(1+\frac{x}{9}\right)-\ln4.\]
    Another submanifold is given by the function $\phi(x)$, which is the solution of the equation:
        \begin{equation}\label{eqq}4x^2\phi^3(x)-x^2\phi^2(x)-18x\phi(x)+4x+27=\frac{d(1-bc)x-
        bd^2}{c(ad-1)x+ad^2},\end{equation}
    where $a,b,c,d$ are arbitrary real numbers. By direct verification, it can be easily shown that the homomorphism
    of linear fractional transformations into the group of translations is provided by the solution $\Phi$ of linear
    fractional form:
        \[\Phi(x)=\frac{ax+b}{cx+d}.\]
    We rephrase the obtained result as
\begin{prop}\label{pp}There exist additive bingles in $\H$ which satisfy the "generalized coplanarity condition",
    with the 3-orthogonality condition (\ref{compf}), if the following conditions (which are sufficient, but not
    necessary), hold true:\par\smallskip
    1) the pair of vectors $\{\va,\vb\}$ satisfies a relation of the form $w_2=w_2(w_1,w_3)$.\par
    2) the intermediate vector $\vc$ is orthogonal (in the Euclidean sense !) to the vector
        \[\va\bigcirc\vb=(a_2b_3+a_3b_2,a_1b_3+a_3b_1,a_2b_1+a_1b_2)\]
    and to the vector $(\va)^{-1}\equiv (1/a_1,1/a_2,1/a_3)$.\par
    If the pair of vectors $\{\va,\vb\}$ satisfies the condition 1, then the bingle additivity condition is
    equivalent to the existence of a homomorphism $\Phi$ of the discrete subgroup of the group diffeomorphisms of the
    real line defined by the equation (\ref{addfin3}). 
    For the submanifold $w_2=0$ and the submanifold given by a relation of the form $w_2=w_1^2\phi(w_1^3/w_3)$,
    where the function $\phi$ is defined by the equation (\ref{eqq}), the expressions of bingles respectively have
    the forms:
        \[{}^3\varphi_1(\va,\vb)=\ln\left(1+\frac{w_1^3}{9w_3}\right)-\ln4;\quad
        {}^3\varphi_2(\va,\vb)=\frac{aw_1^3+bw_3}{cw_1^3+dw_3}.\]
\end{prop}
    We note that, for the obtained solutions, is characteristic the fact, that the additive bingle exists for couples
    of vectors which satisfy a certain condition (a 5-dimensional hypersurface in the 6-dimensional space, or a
    2-dimensional surface in the 3-dimensional projective space of vector coordinates), while in the Euclidean space
    the additive angle exists for any couple of vectors; as well, in the $\H$ case the additivity holds true for a
    1-parameter family of intermediate vectors while in the Euclidean space the family of intermediate vectors
    is 2-dimensional.
\subsection{The nonlinear "coplanarity condition", for a given additive bingle}
    In this subsection we state the problem of finding the additive bingle in its converse form: assuming that a
    certain expression for bingles is additive, we find the "coplanarity condition", which make the imposed
    additivity condition hold true. In such a formulation of the additive bingle problem, the question of solving
    functional equations of additivity is completely absent.\par
    Let $f(w_1^{ab},w_2^{ab},w_3^{ab})$ be the given bingle function between a couple of vectors, and
    assume satisfied the bingle additivity condition with intermediate vector $\vc$, of the form:
        \begin{equation}\label{Add22b}f(w^{ab}_1,w^{ab}_2,w^{ab}_3)=f(w^{ac}_1,w^{ac}_2,w^{ac}_3)+
        f(w^{cb}_1,w^{cb}_2,w^{cb}_3).\end{equation}
    We need to find the relation $\vc=\vc(\va,\vb)$, which makes the equality (\ref{Add22b}) hold true. Written w.r.t.
    components, this dependence leads to three functions of six variables. A formal approach to this problem shows
    that the number of solutions is infinite. Indeed, if we fix two arbitrary functions $c^i$ of three possible
    ones, the equation (\ref{Add22b}) can be regarded as an equation in terms of the remaining third function, which,
    dropping the degenerate cases, always admits (possibly implicit) solutions. Obviously, of special interest are
    the "coplanarity conditions", which satisfy the set of basic properties: homogeneity, covariance, symmetry, etc.
    An example of such relations are already pointed out in the previous subsections of this section.\par
    As nontrivial example, we investigate a slight generalization of the coplanarity relation, described in section
    \ref{nontr}. We formulate the result as a statement, whose validity can be straightforward verified.
\begin{prop}\label{pc}The bingle of the form
        \begin{equation}\label{bingnonl}{}^3\varphi(\va,\vb)=\ln[(w_1^{ab})^A(w_2^{ab})^B(w_3^{ab})^C]+\ln D,
        \end{equation}
    where $A,B,C,D$ are arbitrary real constants $(D>0)$, becomes additive only if the "coplanarity condition" given
    by the following system of equations, is fulfilled:
        \[w_1^{ac}w_1^{cb}=\gamma^B D^{-1/A}(w_1^{ab})^{1-\beta B/A}(w_2^{ab})^\alpha(w_3^{ab})^{-kB};\quad
        w_2^{ac}w_2^{cb}=\gamma^{-A}(w_1^{ab})^{\beta}(w_2^{ab})^{1-\alpha A/B}(w_3^{ab})^{kA},\]
    where $\alpha,\beta,\gamma$ are arbitrary real numbers. The first of the "coplanarity conditions" can be ignored,
    if $A=0$, the second --- for $B=0$; moreover, if $A=B=0$, then both equations can be ignored, and the bingle
    (\ref{bingnonl}) is unconditionally additive.
\end{prop}
    In the case $A\cdot B\neq0$, the manifold of admissible intermediate vectors is 1-dimensional for $AB=0$,
    and 2-dimensional, for $A^2+B^2\neq0$.\par
    We shall further examine as example, the case: $A=1$, $B=C=\alpha=0$, $D=1/3$. Here, the explicit coplanarity
    condition has the form:
        \[\left(\frac{c_1}{a_1}+\frac{c_2}{a_2}+\frac{c_3}{a_3}\right)\left(\frac{b_1}{c_1}+\frac{b_2}{c_2}+
        \frac{b_3}{c_3}\right)=3\left(\frac{b_1}{a_1}+\frac{b_2}{a_2}+\frac{b_3}{a_3}\right)\]
    or, denoting $x_i=c_i/a_i$, $\xi_i=b_i/a_i$, this becomes:
        \[(x_1+x_2+x_3)\left(\frac{\xi_1}{x_1}+\frac{\xi_2}{x_2}+\frac{\xi_3}{x_3}\right)=3(\xi_1+\xi_2+\xi_3).\]
    The last equation provides in the domain $x_i>0$ a conic surface, which is generated by rotating in a certain
    manner, a ray which emerges from the point $x_i=0$. Using the parametric equations
        \[x_1=\frac{1}{3}+u+v;\quad x_2=\frac{1}{3}-u;\quad x_3=\frac{1}{3}-v,\]
    of the plane $x_1+x_2+x_3=1$, we obtain the equation of the section of the conic surface provided by the plane:
        \[\frac{(u+v)\xi_1}{1/3+u+v}+\frac{u\xi_2}{u-1/3}+\frac{v\xi_3}{v-1/3}=0.\]
    This resulting curve is of 4-th order. For $\xi_1=\xi_2=\xi_3$, its shape is represented in the figure
    \ref{1}.\par
        \begin{center}{\small\refstepcounter{figure}\label{1}
        \includegraphics[width=.5\textwidth, height=0.5\textwidth]{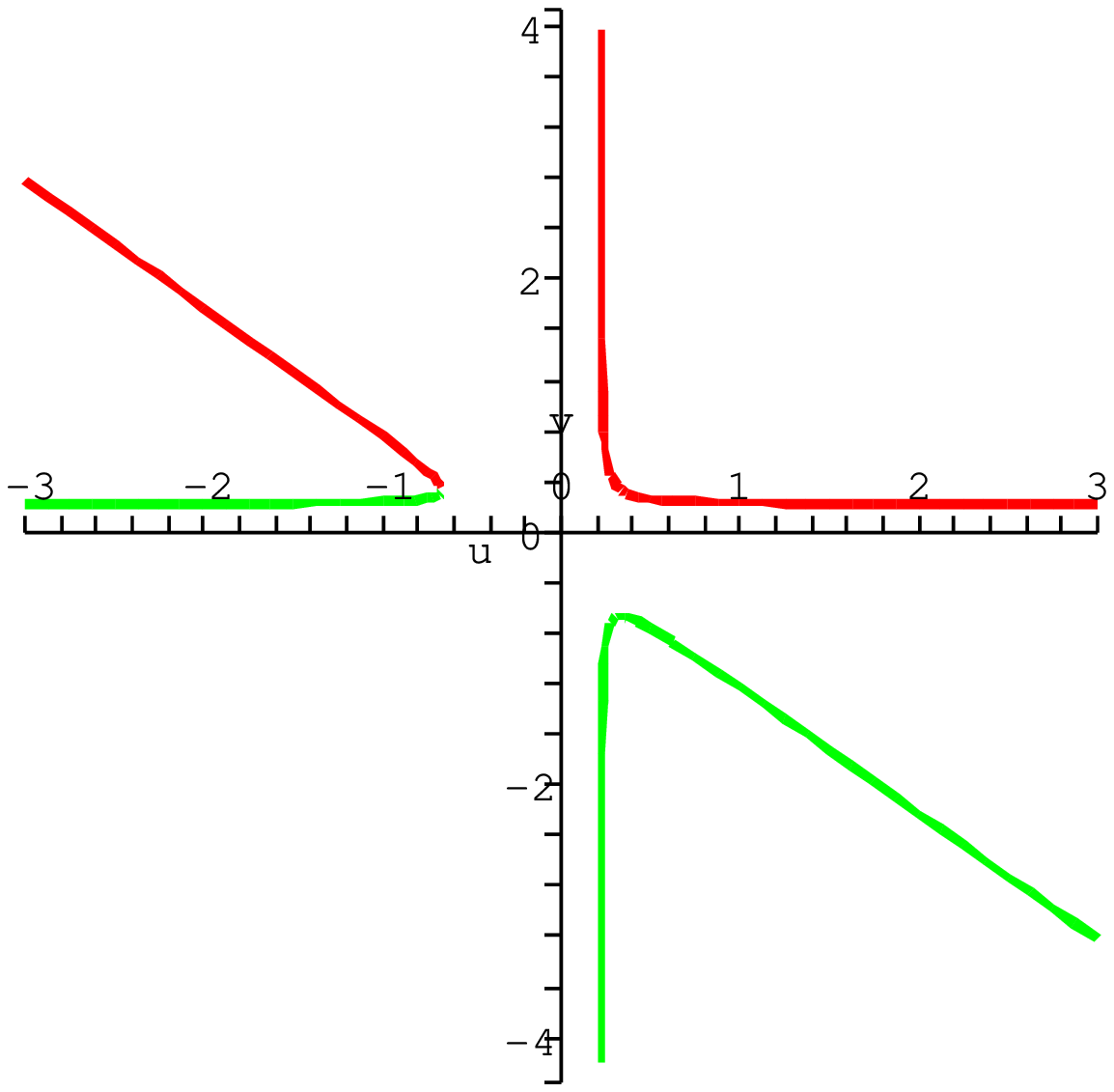}\medskip\nopagebreak\par
        Fig.\thefigure.
        The intersection of the coplanarity surface with the plane $x_1+x_2+x_3=1$. The origin of the coordinate
        system $(v,u)$ coincides with the intersection point of this surface with the straight line
        $x_1=x_2=x_3$.}\end{center}
\section{Additive tringles in $\H$.}\label{tringles}
    The conformally invariant tringle $\varphi(\va,\vb,\vc)$, built on three vectors depends only on the
    following 9 independent conformal invariants:
        \[w_1^{ab}=\xi_1+\xi_2+\xi_3;\quad w_1^{ac}=\eta_1+\eta_2+\eta_3;\quad
        w_1^{bc}=\frac{\eta_1}{\xi_1}+\frac{\eta_2}{\xi_2}+\frac{\eta_3}{\xi_3};\]
        \[w_2^{ab}=\xi_1\xi_2+\xi_2\xi_3+\xi_1\xi_3;\quad w_2^{ac}=\eta_1\eta_2+\eta_2\eta_3+\eta_1\eta_3;\quad
        w_2^{bc}=\frac{\eta_1\eta_2}{\xi_1\xi_2}+\frac{\eta_2\eta_3}{\xi_2\xi_3}+\frac{\eta_1\eta_3}{\xi_1\xi_3};\]
        \[w_3^{ab}=\xi_1\xi_2\xi_3;\quad w_3^{ac}=\eta_1\eta_2\eta_3;\quad\]
        \[w_4^{abc}=\frac{{}^3G(\va,\vb,\vc)}{{}^3G(\va,\va,\va
        )}\xi_2\eta_3+\xi_3\eta_2+\xi_1\eta_3+\xi_3\eta_1+\xi_1\eta_2+\xi_2\eta_1.\]
    Then, in the general case we have:
        \[\varphi(\va,\vb,\vc)=f(w_1^{ab},w_1^{ac},w_1^{bc},w_2^{ab}w_2^{ac},w_2^{bc},w_3^{ab},
        w_3^{ac},w_4^{abc}).\]
    The form of the additivity condition for tringles is not as evident as for the bingle case, since there exist no
    analogues for it in the Euclidean geometry\footnote{We note, that the geometry of angles for rigid bodies from
    3D-mechanics reduces to the geometry of angles in plane, and hence this cannot be regarded as an Euclidean
    variant of the geometry of tringles.}. Without pretending to extensively exhaust the problem, we shall describe
    several relatively simple and symmetric alternatives for the condition of additivity of tringles, where $\vd$ is
    the 4-th (intermediate) vector:
        \begin{enumerate}\item
        $\varphi(\va,\vb,\vc)=\varphi(\vd,\vb, \vc)+\varphi(\va,\vd, \vc)+\varphi(\va,\vb, \vd);$\item
        $\varphi(\va,\vb, \vc)=\varphi(\va,\vb, \vd)+\varphi(\vb,\vc, \vd)+\varphi(\vc,\va, \vd);$\item
        $\varphi(\va,\vb, \vc)=\sum\limits_\nu(-1)^{\sigma_\nu}\varphi(S_\nu(\widehat{\va,\vb,\vc,}\vd))$,
    where $S_\nu$ represents the sum of all permutations of the three vectors, which can be selected from the set of
    four vectors $(\va,\vb,\vc,\vd)$ by means of omitting one of the {\em first three} vectors (the hat
    $\widehat\;$ denotes the corresponding omission), and the signature $\sigma_\nu$ of the permutation $S_\nu$
    is equal to the signature of the permutation of vectors in the arguments of $\varphi$ relative to the initial
    permutation of $(\va,\vb,\vc)$, if the vector $\vd$ is replaced with the missing third vector of the initial
    triple in this set of arguments.\item
    $\varphi(\va,\vb, \vc)=\varphi(\va,\vd, \vd)+\varphi(\vd, \vb, \vd)+\varphi(\vd, \vd, \vc)$.\end{enumerate}
    We can immediately remark, that the last variant of the additivity condition admits a simple extension of the
    naturally-invariant bingle (\ref{solg3}):
        \[{}^3\varphi_{(A,B)}(\va,\vb,\vc)=\ln((w_3^{ab})^A(w_3^{ac})^B),\]
    which has the following properties:
        \[\begin{array}{l}
        {}^3\varphi_{(A,B)}(\va,\va,\va)=0;\quad {}^3\varphi_{(A,B)}(\va,\va,\vc)=B{}^3\varphi(\va,\vc);
        \medskip\\
        {}^3\varphi_{(A,B)}(\va,\vc,\vb)={}^3\varphi_{(B,A)}(\va,\vb,\vc);\quad
        {}^3\varphi_{(A,B)}(\vb,\va,\vc)={}^3\varphi_{(-A-B,B)}(\va,\vb,\vc).\end{array}\]
    Having in view the additivity relations considered above we shall further call "generalized additivity", the
    ability of a tringle to be expressed (in some manner) in terms of tringles built on other triples of
    vectors which are (or not) related to the initial triple. We have in view relations of the form:
        \begin{equation}\label{addgen}\begin{array}{ll}
        {}^3\varphi(\va,\vb,\vc)=&f({}^3\varphi(\vec s_{11},{}^3\vec s_{12},
        \vec s_{13}),\dots,{}^3\varphi(\vec s_{i1},\vec s_{i2},\vec s_{i3}),{}^3\medskip\\
        &\dots,{}^3\varphi(\vec s_{n1},\vec s_{n2},\vec s_{n3}),{}^2\varphi(\vec t_{11},\vec t_{12}),\medskip\\
        &\dots,{}^2\varphi(\vec t_{j1},\vec t_{j2}),\dots,{}^2\varphi(\vec t_{n1},\vec t_{n2})),
        \end{array}\end{equation}
    where at least one of the vectors $\vec s_{ij}$ and $\vec t_{kl}$ for each $i$ and $k$ does not belong to the
    triple $\va,\vb,\vc$.\par
    It is clear, that in this setting the problem of finding "additive tringles" has infinitely many solutions, which
    almost always will be implicitly defined by means of huge algebraic expressions.
\subsection{Dimensionless expressions for the basic conformal \protect{\\}invariants}
    In this section we provide for future reference the full range of basic conformal invariants in dimensionless form.
    Like in the previous sections, we shall use the following dimensionless notations:
        \[\frac{b_i}{a_i}=\xi_i;\quad \frac{c_i}{a_i}=\eta_i;\quad\frac{d_i}{a_i}=\delta_i.\]
   Moreover, for simplifying the formulas, we introduce as well the notations:
        \[\Delta_p=p_1p_2p_3;\quad \Delta_1(p,q)=p_2q_3+p_3q_2;\quad \Delta_2(p,q)=p_1q_3+q_3p_1;\quad
        \Delta_3(p,q)=p_1q_2+p_2q_1;\]
        \[w_1(p)=p_1+p_2+p_3;\quad w_2(p)=p_1p_2+p_2p_3+p_1p_3,\]
   for all the triples $\{p_1,p_2,p_3\}$, $\{q_1,q_2,q_3\}$. We note that $p/q\equiv\{p_1/q_1,p_2/q_2,p_3/q_3\}$.
   The calculation of the basic conformal invariants listed at the beginning of Section \ref{tringles}, for all
   the possible combinations of triples and couples of vectors from the set $\{\va,\vb,\vc,\vd\}$, leads to the
   following system of expressions:
        \[6w_4^{abc}=\Delta_1(\xi,\eta)+\Delta_2(\xi,\eta)+\Delta_3(\xi,\eta);\]
        \[6w_4^{abd}=w_4^{adb}=\Delta_1(\xi,\delta)+\Delta_2(\xi,\delta)+\Delta_3(\xi,\delta);\quad\]
        \[6w_4^{acd}=6w_4^{adc}=\Delta_1(\eta,\delta)+\Delta_2(\eta,\delta)+\Delta_3(\eta,\delta);\quad\]
        \[w_4^{bad}=w_4^{bda}=w_4^{abd}/\Delta_\xi;\quad w_4^{cad}=w_4^{cda}=w_4^{acd}/\Delta_\eta;\]
        \[6w_4^{bcd}=6w_4^{bdc}=\Delta_1(\eta/\xi,\delta/\xi)+\Delta_2(\eta/\xi,\delta/\xi)+\Delta_3(\eta/\xi,
        \delta/\xi);\]
        \[6w_4^{cbd}=6w_4^{cdb}=\Delta_1(\xi/\eta,\delta/\eta)+\Delta_2(\xi/\eta,\delta/\eta)+\Delta_3(\xi/\eta,
        \delta/\eta);\]
        \[w_4^{dab}=w_4^{dba}=w_4^{abd}/\Delta_\delta; w_4^{dac}=w_4^{dca}=w_4^{acd}/\Delta_\delta;\]
        \[6w_4^{dbc}=6w_4^{dcb}=\Delta_1(\eta/\delta,\xi/\delta)+\Delta_2(\eta/\delta,\xi/\delta)+
        \Delta_3(\eta/\delta,\xi/\delta);\]
        \[w_3^{ab}=\frac{1}{w_3^{ba}}=\Delta_\xi;\quad w_3^{ac}=\frac{1}{w_3^{ca}}=\Delta_\eta;\quad
        w_3^{ad}=\frac{1}{w_3^{da}}=\Delta_\delta;\]
        \[w_3^{bc}=\frac{1}{w_3^{cb}}=\frac{\Delta_\eta}{\Delta_\xi};\quad
        w_3^{bd}=\frac{1}{w_3^{db}}=\frac{\Delta_\delta}{\Delta_\xi};\quad
        w_3^{cd}=\frac{1}{w_3^{dc}}=\frac{\Delta_\delta}{\Delta_\eta};\]
        \[3w_2^{ab}=w_2(\xi);\quad 3w_2^{ac}=w_2(\eta);\quad 3w_2^{ad}=w_2(\delta);\quad\]
        \[3w_2^{ba}=w_1(\xi)/\Delta_\xi;\quad 3w_2^{ca}=w_1(\eta)/\Delta_\eta;\quad
        3w_2^{da}=w_1(\delta)/\Delta_\delta;\]
        \[3w_2^{bc}=w_2(\eta/\xi);\quad 3w_2^{cb}=w_2(\xi/\eta);\quad 3w_2^{bd}=w_2(\delta/\xi);\quad
        3w_2^{db}=w_2(\xi/\delta);\quad\]
        \[3w_2^{cd}=w_2(\delta/\eta);\quad 3w_2^{dc}=w_2(\eta/\delta).\]
    From the presented invariants, only 21 are functionally independent. The essence of building generalized-additive
    tringles relies on the choice out of the 21 invariants, of a subset of not less than 9 invariants, solvable in
    terms of 9 variables --- the components of $\{\xi,\eta,\delta\}$. The substitution of the solvable equations in any other
    independent invariant provides a formula of the form (\ref{addgen}).
\subsection{Example: the generalized-additive tringle}
    We shall examine the following relations, which emerge from the systems for invariants of the preceding section:
        \[w_1(\xi)=3w_2^{ba}\Delta_\xi=3w_2^{ba}\frac{w_4^{abd}}{w_4^{bad}};\quad w_2(\xi)=3w_2^{ab};\quad
        \Delta_\xi=w_3^{ab}w_3^{ba};\]
        \[w_1(\eta)=3w_2^{ca}\Delta_\eta=3w_2^{ca}\frac{w_4^{acd}}{w_4^{cad}};\quad w_2(\eta)=3w_2^{ac};\quad
        \Delta_\eta=w_3^{ac}w_3^{ca};\]
        \[w_1(\delta)=3w_2^{da}\Delta_\delta=3w_2^{da}\frac{w_4^{acd}}{w_4^{dca}};\quad w_2(\delta)=3w_2^{da};\quad
        \Delta_\xi=w_3^{ad}w_3^{da}.\]
    These expressions contain nine symmetric polynomials in the variables $\{\xi_i\}$, $\{\eta_i\}$ and
    $\{\delta_i\}$, expressend by means of the the invariants. The procedure of solving the systems of symmetric
    equations in terms of these variables, having in view the relations between symmetric polynomials and the roots
    of algebraic equations, can be resumed to finding triples of real solutions for the triple of cubic equations of
    the form:
        \begin{equation}\label{cub1}\xi^3-3w_2^{ba}\frac{w_4^{abd}}{w_4^{bad}}\xi^2+3w_2^{ab}\xi-
        w_3^{ab}w_3^{ba}=0;\end{equation}
        \begin{equation}\label{cub2}\eta^3-3w_2^{ca}\frac{w_4^{acd}}{w_4^{cad}}\eta^2+
        3w_2^{ac}\eta-w_3^{ac}w_3^{ca}=0;\end{equation}
        \[\delta^3-3w_2^{da}\frac{w_4^{acd}}{w_4^{dca}}\delta^2+3w_2^{da}\delta-w_3^{ad}w_3^{da}=0.\]
    We shall further examine as expression for the generalized additive tringle the invariant $w_4^{abc}$, or some
    function depending on this invariant. The condition of generalized  additivity has the form:
        \[w_4^{abc}=\Delta_1(\xi,\eta)+\Delta_2(\xi,\eta)+\Delta_3(\xi,\eta)|^{\xi_i=\bar\xi_i(w_2^{ba}
        w_4^{abd}/w_4^{bad},w_2^{ab},w_3^{ab}w_3^{ba})}_{\eta_i=\bar\eta_i(w_2^{ca}w_4^{acd}/w_4^{cad},
        w_2^{ac},w_3^{ad}w_3^{da})}\]
    which represents a quite massive algebraic expression, depending on the shown combinations of conformal
    invariants. Here $\bar\xi_i$ and  $\bar\eta_i$ are the triples of roots of the cubic equations
    (\ref{cub1})-(\ref{cub2}), which are expressed in terms of the coefficients of these equations.
\section{Exponential angles}
    In \cite{pavlov}, it was shown that the hypercomplex numbers from $\mathcal{H}_n$ admit an exponential
    representation. In this section we shall derive the expressions of exponential angles and determine their
    relation to the problem stated in the present paper. We first recollect the main data, which are needed for
    obtaining the formulas for exponential angles.\par
    The hypercomplex numbers - as elements of the algebra $\mathcal{H}_n$ will be represented as linear
    combinations:
    \[a=a_1i_1+\dots a_ni_n,\]
    where $i_k$ are the generators of the algebra $\mathcal{H}_n$, which satisfy the following multiplication rule:
    $i_k\cdot i_l=i_k\delta_{lk}$ (the isotropic system of generators). The norm of an element $a$ is the real number
    $|a|=|a_1\cdot\dots\cdot a_n|^{1/n}$. The operations of multiplication and division of two elements
    $a=a_1i_1+\dots a_ni_n$, $b=b_1i_1+\dots b_n i_n$ are given by:
        \[a\cdot b=a_1b_1i_1+\dots+a_nb_ni_n;\quad a/b=(a_1/b_1)i_1+\dots+(a_n/b_n)i_n,\]
    where the division can be performed only w.r.t. elements with nonvanishing norm ($|b|\neq0$).\par
    We call {\em function of hypercomplex variable} $f(a)$, a hypercomplex number of the form:
        \begin{equation}\label{func}f(a)=f(a_1)i_1+\dots f(a_n)i_n.\end{equation}
    If the function $f(x)$ is analytic for real argument $x$, then our definition is automatically satisfied, due
    both to the representation of the function as formal series and to the multiplication table of the algebra
    $\mathcal{H}_n$.\par
    As it has been shown in \cite{pavlov}, any hypercomplex number $a$ having all its components positive ($a_i>0$)
    can be represented in exponential form as:
        \[a=|a|\exp(\alpha_1e_1+\dots+\alpha_{n-1}e_{n-1}),\]
    where $\{\alpha_i\}$ are the exponential angles, $\{e_i\}$ is the special basis of the algebra of the group of
    hyperbolic rotations, which is related to the isotropic basis by the relations:
        \[e_1=i_1-i_2;\quad e_2=i_1-i_3;\quad\dots e_{n-1}=i_1-i_n.\]
    We examine further the case of the algebra $\H$. In this algebra, the general formulas for the exponential
    representation have the concrete form:
        \[a=a_1i_1+a_2i_2+a_3i_3=(a_1a_2a_3)^{1/3}\exp((\alpha_1+\alpha_2)i_1-\alpha_1i_2-\alpha_1i_3).\]
    Using in the left hand side of the equality the formula (\ref{func}) and identifying the components of the
    hypernumbers from both sides of the equality, we infer the expressions of angles in terms of isotropic
    coordinates:
        \begin{equation}\label{eangle}\alpha_1=\frac{1}{3}\ln\left(\frac{a_1a_3}{a_2^2}\right);\quad
        \alpha_2=\frac{1}{3}\ln\left(\frac{a_1a_2}{a_3^2}\right).\end{equation}
    It is then natural to express the angles $\phi_1(a,b)$ and $\phi_2(a,b)$ formed by the vectors-hypernumbers $a$
    and  $b$ by means of the formulas:
        \[\frac{|a|}{|b|}\exp(\phi_1(a,b)(i_1-i_2)+\phi_2(a,b)(i_1-i_3))=a/b.\]
    We obtain from (\ref{eangle}) the explicit formulas for these angles, after replacing $a_i\to a_i/b_i$:
        \begin{equation}\label{eangle1}\phi_1(a,b)=\frac{1}{3}\ln\left(\frac{a_1a_3b_2^2}{b_1b_3a_2^2}\right);\quad
        \phi_2(a,b)=\frac{1}{3}\ln\left(\frac{a_1a_2b_3^2}{b_1b_2a_3^2}\right).\end{equation}
    It is straightforward to remark that each of the angles is conformally-invariant, is additive in the usual sense
    and vanishes for equal vectors-hypernumbers. However, these angles (whch in our terminology, are {\em bingles})
    will not represent the subject of our research, since they are not expressed in terms of metric invariants
    built exclusively using the vectors-hypernumbers $a$ and $b$. Indeed, it is easy to check that the representation
    (\ref{eangle}) is equivalent to the one which is expressed in terms of metric invariants:
        \[\phi_1(a,b)=\frac{1}{3}\ln\left(\frac{(a,a,i_2)(i_1,i_3,b)^2}{(b,b,i_2)(i_1,i_3,a)^2}\right);\quad
        \phi_2(a,b)=\frac{1}{3}\ln\left(\frac{(a,a,i_3)(i_1,i_2,b)^2}{(b,b,i_3)(i_1,i_2,a)^2}\right).\]
    We can see that {\em the exponential angles between vectors-hypernumbers (bingles) are defined by means of
    a third set of vectors --- the isotropic basis in $\H$.} This situation is clearly different from the one
    considered in the setting of the problem investigated in our paper, since this setting emerges from Euclidean
    geometry. {\em The existence of the fundamental invariant isotropic directions in $\mathcal{H}_n$ gives the
    possibility to generalize the notion and the definition of angle in such a way, that the presence of these
    directions is explicitly involved.} It is clear then, that in the case Euclidean geometry, in view of absence of
     any invariant directions, such definitions cannot be formulated.

\bigskip

We are thankful to Professor V. Balan for fruitful discussion and
translation assistance.


\begin{thebibliography}{1}
\bibitem{pavlov}G.I. Garasko and D.G. Pavlov, {\em The geometry of non-degenerate polynumbers} (in Russian),
    Hypercomplex Numbers in Geometry and Physics, vol. 4, no. 1 (7) (2007), 3-25.
\end{thebibliography}
\end{document}